# Study of SiGeSn/GeSn single quantum well towards high-performance all-group-IV optoelectronics

Grey Abernathy,[1,2] Yiyin Zhou,[1,2] Solomon Ojo,[1,2] Bader Alharthi,[1] Perry C. Grant,[1,2] Wei Du,[3*] Joe Margetis,[4] John Tolle,[4] Andrian Kuchuk,[5] Baohua Li,[6] and Shui-Qing Yu[1,5*]

[1]Department of Electrical Engineering, University of Arkansas, Fayetteville, Arkansas 72701, USA

[2]Microelectronics-Photonics Program, University of Arkansas, Fayetteville, Arkansas 72701, USA

[3]Department of Electrical Engineering and Physics, Wilkes University, Wilkes Barre, Pennsylvania 18766, USA

[4]School of Electrical, Energy and Computer Engineering, Arizona State University, Tempe, Arizona 85287, USA

[5]Institute for Nanoscience and Engineering, University of Arkansas, Fayetteville, Arkansas 72701, USA

[6]Arktonics, LLC, 1339 South Pinnacle Drive, Fayetteville, Arkansas 72701, USA

The recent progress on (Si)GeSn optoelectronic devices holds a great promising for photonic integration on the Si substrate. In parallel to the development of bulk devices, the (Si)GeSn based quantum wells (QWs) have been investigated aiming to improve the device performance. While the multiple QW structure is preferred for the device application, the single quantum well (SQW) is more suitable for optical property study. In this work, a comprehensive study of a SiGeSn/GeSn SQW was conducted. The calculated band diagram provided the band alignment and energies of possible transitions. The SQW features the direct bandgap well with L-Γ valley energy separation of 50 meV, and the barrier heights for both electron and hole are greater than 80 meV. Using two continuous-wave and two pulsed pumping lasers, the analysis of PL spectra allows for identifying different transitions and for better understanding the SQW optical properties. The study could provide the guidance for advancing the future QW design towards device applications.

Keywords: Silicon germanium tin, germanium tin, quantum well, optical transition

* PoC Email: syu@uark.edu; wei.du@wilkes.edu

## I. INTRODUCTION

Lasers that are capable of monolithically integrated on Si have been highly desirable.[1,2] For the past decades, owing to the bandgap directness nature of group III-V materials, they have been widely utilized in Si photonics via hybrid integration such as direct growth on Si or wafer bonding, making them dominated approaches.[3-5] It was well acknowledged that group-IV materials such as Si, Ge, and SiGe alloys feature indirect bandgap and therefore they suffer from low efficiency of optical transitions. However, recent study on group-IV alloy GeSn opens a new avenue for Si-based lasers. Theoretical studies have indicated that GeSn could possess direct bandgap with sufficient Sn composition of 6~10%.[6-8] Thanks to the cutting-edge material growth technology, growth of GeSn alloy with Sn incorporation beyond the indirect-to-direct bandgap transition point is attainable despite the low solid solubility of Sn in Ge (~1%).[9-12]

A breakthrough of true direct bandgap GeSn being experimentally identified was reported in 2014, followed by the first demonstration of the optically pumped GeSn lasers in 2015.[13,14] The promising results inspire the community to further develop the GeSn technology, and since then there was considerable progress made on GeSn-based lasers, including: i) the lasing spectrum covers 2-3 µm showing the tunable emission wavelength;[15] ii) the Sn composition in GeSn active layer exceeded 22%, while it is less than 13% in the first GeSn laser;[16] iii) the lasing operational temperature is near room temperature under pulsed pumping;[17,18] iv) the continuous-wave (CW) operation was achieved at low temperature up to 70 K;[19] and v) the first electrically injected edge-emitting laser was demonstrated and the lasing temperature is up to 100 K.[20]

Following the current bulk GeSn laser development, the low threshold laser using quantum well (QW) structure would naturally be the next step. For the past five years, various GeSn QWs have been grown and characterized, aiming to deliver the viable structures for high-performance lasers.

To achieve this goal, it is well acknowledged that the direct bandgap QW with type-I band alignment is preferred. The summary of relevant QW studies is shown in Table I.[21-31]

Using relaxed Ge as buffer layer was reported in early QW studies.[21,22,27] Since the barriers and well are generally grown pseudomorphically on Ge buffer, the QW layer is under relatively large compressive strain and will yield the indirect bandgap even the alloys with high Sn composition were used. In refs. 21-22, it can be seen that even though the wider bandgap materials such as Ge, GeSn (with lower Sn composition), and SiGeSn were used as barriers, the indirect bandgap well deteriorates the emission efficiency despite of the large barrier heights in both conduction band (CB) and valance band (VB).

To ease the compressive strain in QW, the use of a relaxed GeSn buffer in addition to a Ge buffer prior to the growth of QW was proposed and implemented, which promotes the bandgap directness of the well. Moreover, the GeSn buffer with the lower Sn composition also serves as a barrier relative to higher Sn composition well. A 14% Sn well associated with 8% Sn buffer (barrier) was reported, showing the direct bandgap with $\Gamma$-L separation of 22 meV and the barrier height of 20 meV in $\Gamma$-valley.[24] A follow-up study using higher Sn compositions of 15.3% and 9.4% in both well and buffer (barrier) exhibited the improved bandgap directness ($\Gamma$-L separation of 52 meV) and enhanced carrier confinement (barrier height of 35 meV in $\Gamma$-valley).[25] However, the limit of this approach is that the Sn composition difference between GeSn buffer and well cannot be too large, as reported of ~6%, otherwise the increased compressive strain in the well would neither further facilitate the bandgap directness nor improve the barrier height. Another issue is that with higher than 8% Sn the barrier layer features direct bandgap as well, which might enhance the optical transitions in barrier and thus demote the QW emission.

Although the GeSn buffer addressed the direct bandgap issue in the well, using it also as barrier could not provide the sufficient carrier confinement particularly at higher temperature, which hampers the device room temperature operation. Consequently, using SiGeSn as barrier above the GeSn buffer was attempted, being motivated by the theory that if the Si and Sn compositions were appropriately selected, a wider bandgap layer with sufficient barrier height is available, while the lattice constant can also be individually engineered to reduce the strain effect.[28,29] The preliminary study on SiGeSn/GeSn/SiGeSn QW structures led to the demonstration of GeSn QW lasers.[26,30] The multi-quantum-well (MQW) configurations were used to enhance the modal gain. Indeed, the significantly reduced lasing threshold compared with that of bulk laser was obtained. On the other hand, the maximum lasing temperature was reported as 90 K, indicating that there is considerable room to improve the laser performance by optimizing the QW design.

Based on the published results that are shown in Table I, some consensuses regarding QW design can be summarized as following: i) besides Ge buffer, growing a relaxed GeSn buffer with Sn composition lower than that in well could ease the compressive strain and therefore facilitate the direct bandgap well; ii) for the barrier, SiGeSn could provide larger barrier heights at both CB and VB, and therefore offer better carrier confinement than using GeSn. However, with current material growth capability the choice of Si and Sn compositions is limited; iii) MQW is preferred for device applications, and growth of identical wells is critical. If a design requires several pairs QW, the precise control on material growth is a challenge; and iv) there are issues regarding band alignment. The wider bandgap SiGeSn barrier is most likely under tensile strain, which lifts its light hole (LH) band above the heavy hole (HH) band. Considering the compressive strain in well that brings down the LH band, there is no hole confinement at LH band.

Table I Summary of QW study[21-31]

| Buffer | SQW / MQW | Bottom barrier / Top barrier | Sn% in QW | L-Γ valley separation | Γ valley confinement | HH band confinement |
|---|---|---|---|---|---|---|
| Ge | SQW | Ge$_{0.95}$Sn$_{0.05}$ | 9.0% | -20 meV* | 45 meV | 35 meV |
| Ge | SQW | Si$_{0.12}$Ge$_{0.79}$Sn$_{0.09}$ | 9.5% | -10 meV* | 240 meV | 50 meV |
| Ge | MQW | Ge | 8.0% | -50 meV* | 30 meV | 140 meV |
| Ge | SQW | Si$_{0.120}$Ge$_{0.795}$Sn$_{0.085}$ / Si$_{0.120}$Ge$_{0.787}$Sn$_{0.093}$ | 9.5% | -10 meV* | 200 meV | 60 meV |
| Ge$_{0.914}$Sn$_{0.086}$ | MQW | Si$_{0.1}$Ge$_{0.8}$Sn$_{0.1}$ | 8.5% | -5 meV* | 10 meV | 45 meV |
| Ge$_{0.900}$Sn$_{0.100}$ | MQW | Si$_{0.048}$Ge$_{0.822}$Sn$_{0.130}$ | 13.3% | 40 meV | 20 meV | N. A. |
| Ge$_{0.900}$Sn$_{0.100}$ | MQW | Si$_{0.052}$Ge$_{0.814}$Sn$_{0.134}$ | 13.5% | 30 meV | 15 meV | N. A. |
| Ge$_{0.920}$Sn$_{0.080}$ | SQW | Ge$_{0.92}$Sn$_{0.08}$ | 14.0% | 22 meV | 20 meV | 55 meV |
| Ge$_{0.912}$Sn$_{0.088}$ | DQW | Ge$_{0.912}$Sn$_{0.088}$ | 14.7% | 23 meV | 17 meV | 52 meV |
| Ge$_{0.906}$Sn$_{0.094}$ | DQW | Ge$_{0.906}$Sn$_{0.094}$ | 15.3% | 52 meV | 35 meV | 52 meV |
| Ge$_{0.916}$Sn$_{0.084}$ | MQW | Si$_{0.050}$Ge$_{0.887}$Sn$_{0.063}$ | 13.8% | 80 meV | 70 meV | 70 meV |
| Ge$_{0.912}$Sn$_{0.088}$ | MQW | Si$_{0.033}$Ge$_{0.887}$Sn$_{0.080}$ | 14.4% | 90 meV | 75 meV | 75 meV |
| *Ge$_{0.914}$Sn$_{0.086}$ (This work)* | *SQW* | *Si$_{0.042}$Ge$_{0.892}$Sn$_{0.066}$ / Si$_{0.036}$Ge$_{0.897}$Sn$_{0.067}$* | *13.0%* | *50 meV* | *88 meV* | *93 meV* |

* Negative values indicate the indirect bandgap well

Although MQW is more suitable for device, it is difficult to investigate the fundamental optical properties due to the inhomogeneous injection. Therefore, it is more desirable to perform a comprehensive study on single-QW (SQW), including pinpointing the optical transitions, exploring the recombination mechanism, and modeling the band diagram to provide the clear guidance for the future QW design. In this work, a SiGeSn/GeSn/SiGeSn SQW was grown and characterized. Four excitation lasers, two CW and two pulsed with wavelength ranging from 532 to 1950 nm were used to examine the optical transitions via measured PL spectra. Through the comparison with our previous reported SQW that uses GeSn as barrier, the recombination mechanism was clearly disclosed. The simulated electronic band diagram was also presented to assist the understanding of optical properties.

**II. MATERIAL GROWTH AND CHARACTERIZATION**

The QW sample studied in this work was grown using an industrial standard reduced pressure chemical vapor deposition (RPCVD) reactor with the low-cost commercially available Si, Ge, and Sn precursors similar to our previous report.[11] The complete structure includes a nominal 700-nm-thick Ge buffer, a nominal 600-nm-thick $Ge_{0.914}Sn_{0.086}$ buffer, and a 9-nm-thick $Ge_{0.87}Sn_{0.13}$ well layer sandwiched by a 78-nm-thick $Si_{0.042}Ge_{0.892}Sn_{0.066}$ bottom barrier and a 54-nm-thick $Si_{0.036}Ge_{0.897}Sn_{0.067}$ top barrier. The detailed material growth method has been reported elsewhere.[11,32] The material characterizations were conducted after growth. The Secondary Ion Mass Spectrometry (SIMS), high-resolution X-ray diffraction (HRXRD) 2θ-ω scan and reciprocal space map (RSM) were employed to identify the Si and Sn compositions, the degree of strain, and the layer thickness.

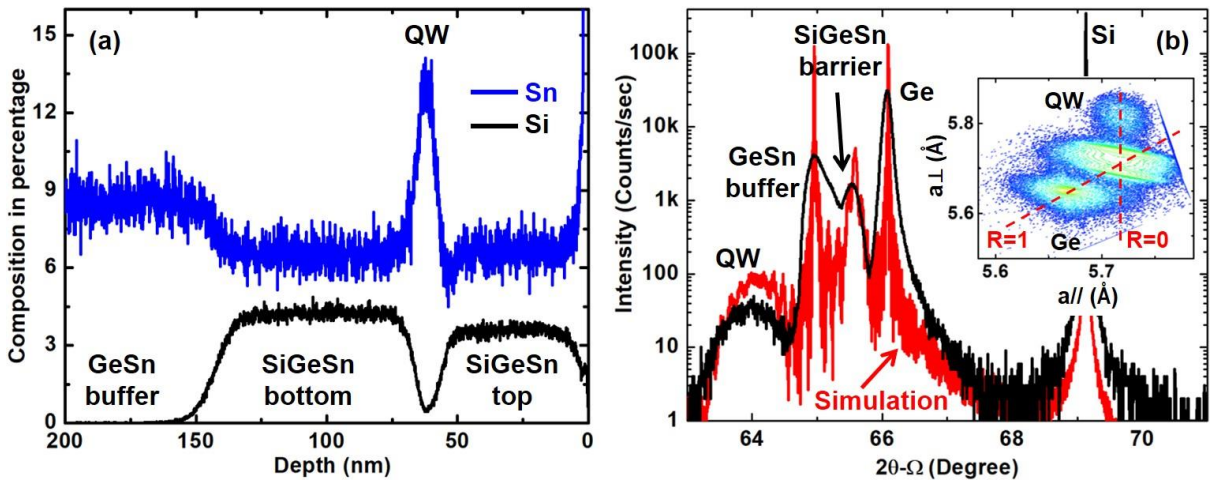

Figure 1 (a) SIMS showing each layer thickness and Si and Sn compositions. (b) XRD 2θ-ω scan showing each resolved peak. The black and red curves are measured data and simulation, respectively. Inset: RSM contour plot.

Figure 1(a) shows the SIMS profile. The Sn compositions in buffer, bottom barrier, well, and top barrier were obtained as 8.6%, 6.6%, 13.0%, and 6.7%, respectively, while the Si compositions in bottom and top barriers were 4.2% and 3.6%. respectively. The thickness of GeSn well can be extracted from the full width at half maximum (FWHM) of SIMS peak.

The HRXRD 2θ-ω scan is shown in Fig. 1(b). The black and red curves are the measured data and simulation results, respectively. The peak at 66.07º corresponds Ge buffer. Generally, the incorporation of Si in Ge would shift the peak towards higher angle degree, while alloying with Sn would shift the peak to the opposite direction. The peaks at 64.02º and 64.94º are assigned to $Ge_{0.87}Sn_{0.13}$ well and $Ge_{0.914}Sn_{0.086}$ buffer, respectively. The peaks indicating SiGeSn bottom and top barrier was observed at 65.54º. Since the Si and Sn compositions are close in bottom and top barriers, their peaks are mostly overlapping. In addition, the XRD simulation was performed, through which the lattice constant and the layer thickness can be determined.

The RSM contour plot is shown in Fig. 1(b) inset. The Ge and GeSn buffers are almost relaxed, with residual tensile and compressive strain of less than 0.2%. The SiGeSn barriers and GeSn well were grown pseudomorphically to GeSn buffer, resulting in the 0.28% tensile strain for SiGeSn barriers and 0.94% compressive strain for GeSn well, respectively. The broaden of SiGeSn contour is due to the slight difference of Si and Sn compositions in bottom and top barriers.

## III. ELECTRONIC BAND STRUCTURE AND OPTICAL TRANSITION ANALYSIS

Using material characterization results as parameters, the electronic band structure was calculated at 300 K and is shown in Fig. 2. The models used for simulation and the band alignment calculation method are described in refs. 16 and 24. It is worth noting that the calculated band structure features non-negligible dependence on bowing parameters. Adoption of various bowing parameters results in spread values in literatures, which indicates that the bowing parameters may be Sn-compositional dependence. In Fig. 2, we use a set of parameters that are consistent with our previous publications and aim to facilitate the analysis of the quantum effect, barrier height, and all possible optical transitions.

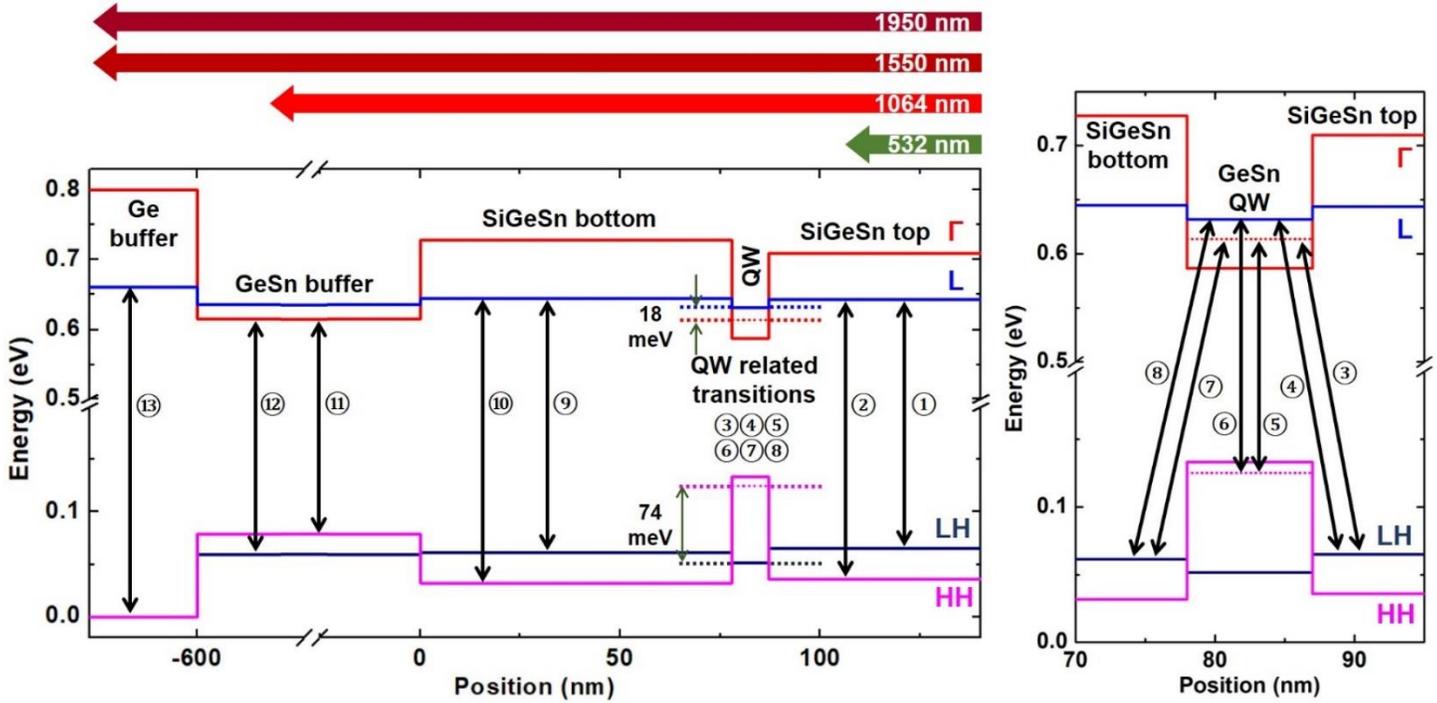

Figure 2 Band diagram and possible optical transitions.

By using $Ge_{0.914}Sn_{0.086}$ buffer in addition to Ge buffer, which mitigates the compressive strain in $Ge_{0.87}Sn_{0.13}$ well, the direct bandgap QW was obtained with the L-Γ valley energy separation of 50 meV. However, due to quantized energy levels in the well, the first energy level in Γ valley is only 18 meV below the L valley minimum; while in VB, the energy separation between the first energy level in HH and LH is 74 meV. In CB, the strong electron confinement is obtained in Γ valley as the barrier heights are greater than 100 meV between GeSn well and SiGeSn barriers. The weak electron confinement of less than 10 meV barrier height in L valley suggests the possible electron leakage channel at high temperature: populating L valley in well from first energy level in Γ valley via thermal excitation, and then scattering to L valleys in SiGeSn barriers. In VB, the barrier heights in HH band are ~ 100 meV, offering excellent hole confinement. However, in LH band, due to the tensile strain of SiGeSn barriers, the LH band sits above HH with a split energy of 40 meV, which leads to a type-II band alignment in LH, and therefore losing hole confinement with respect to well. This may account for the hole leakage channel.

According to the band diagram, the major possible band-to-band optical transitions that could occur in GeSn QW, SiGeSn top and bottom barriers, GeSn buffer, and Ge buffer are labelled with numbers in Fig. 2. The QW related transitions are detailed in the right side. The energy of each transition is listed in Table II, which is organized showing longer to shorter wavelength from top to bottom. Note that for some other possbile transitions, such as from CB in well to HH in barriers, they are not discussed here due to very weak peak intensity which is non-observable.

Table II Summary of possible transitions in each layer
(From top to bottom: longer to shorter wavelength)

|  | **Possible Transitions** | **Energy (meV)** | **Wavelength (nm)** |
|---|---|---|---|
| QW | ⑤ $\Gamma$ – HH | **489** | **2536** |
|  | ⑥ L – HH | **507** | **2446** |
| GeSn buffer | ⑪ $\Gamma$ – HH | 536 | 2313 |
| QW-related (Type II) | ③ $\Gamma$ (QW) – LH (SiGeSn top) | 549 | 2259 |
|  | ⑦ $\Gamma$ (QW) – LH (SiGeSn bottom) | 552 | 2246 |
| GeSn buffer | ⑫ $\Gamma$ – LH | 556 | 2230 |
| QW-related (Type II) | ④ L (QW) – LH (SiGeSn top) | 567 | 2187 |
|  | ⑧ L (QW) – LH (SiGeSn bottom) | 570 | 2175 |
| SiGeSn top | ① L – LH | 579 | 2147 |
| SiGeSn bottom | ⑨ L – LH | 583 | 2127 |
| SiGeSn top | ② L – HH | 608 | 2039 |
| SiGeSn bottom | ⑩ L – HH | 613 | 2023 |
| *Ge buffer* | ⑬ *L – HH/LH* | *660* | *1879* |

## IV. OPTICAL CHARACTERIZATION

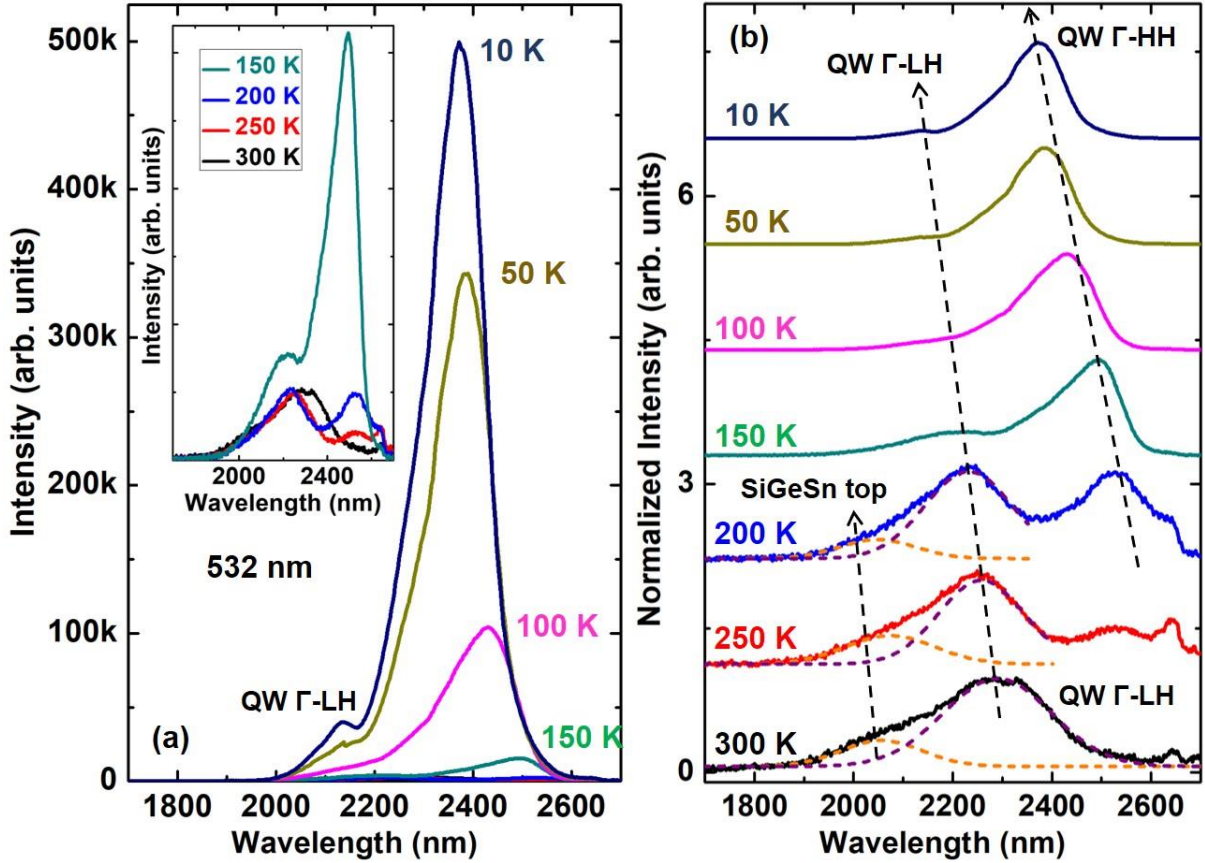

Figure 3 (a) Temperature-dependent PL under 532 nm CW pumping laser. (b) Stacked normalized PL spectra showing the peak fitting.

The comprehensive PL study was performed using the standard off-axis setup configuration and lock-in techniques. To in-depth investigate the optical transition in each layer, two CW and two pulsed lasers were used as pumping sources in this work, including: a 532-nm and a 1550-nm CW lasers, and a 1064-nm and a 1950-nm pulsed lasers. The laser penetration of each laser regarding QW sample structure is shown in Fig. 2 above the band diagram. The emissions were collected using a spectrometer equipped with a PbS detector with spectral cutoff at 3.0 µm.

Figure 3(a) shows the temperature-dependent PL spectra using 532-nm pumping laser. At higher temperature (above 250 K), a broad peak between 2000 to 2400 nm dominates the PL. As temperature decreases to 200 K, a longer wavelength peak at ~2500 nm emerges. As temperature further decreases, both two peaks grow with the intensity of longer wavelength peak increases much more rapidly than that of shorter wavelength peak. Below 200 K, the longer wavelength peak becomes dominate the PL.

Figure 3(b) shows the normalized temperature-dependent PL spectra that were stacked for clarity. At higher temperature, the broad peak between 2000 to 2400 nm could include multiple optical transitions according to transition energies that are listed in Table II. To better identify the origin of the transitions, the broad peak was fitted into two peaks using Gaussian distribution at 300, 250, and 200 K, as shown in Fig. 3(b) bottom. The detailed fitting information is shown in Table III. Note that the penetration depth of 532-nm laser is ~ 50 nm in GeSn/SiGeSn material system, and thus the major absorption occurs in SiGeSn top barrier. At 300 K, the fitted shorter wavelength peak (peak 1) at 2056 nm may be assigned to L valley to LH and L valley to HH transitions in SiGeSn top barrier (① and ②), and these transitions cannot be further identified due to small energy separation. The fitted longer wavelength peak (peak 2) sits at 2290 nm, which is attributed to the transitions from CB in QW to LH in SiGeSn top barrier, i.e., first energy level in Γ valley to LH (③) and L valley to LH (④). Note that the calculated wavelengths of ③ (2259 nm) and ④ (2178 nm) are shorter than fitted wavelength of 2290 nm, which may be due to selected bowing parameters. The QW emission is hardly seen from 300 K PL. This is because of: i) the absorption in QW is relatively weak under 532nm laser pumping; and ii) as a result of 18 meV energy difference between first energy level in Γ valley and L valley in QW, the carrier collection efficiency is low since the carriers in Γ valley could be thermally excited to L valley and then spread to SiGeSn barrier.

As temperature decreases from 300 to 200 K, both fitted peaks feature blue shift as expected. Moreover, a peak at ~2500 nm was observed and its intensity comparable with the shorter wavelength one at 200 K. This peak is assigned to the transition in QW between first energy levels in Γ valley and HH band (⑤), which can be interpreted by the improved carrier confinement at lower temperature, since the carrier lifetime becomes longer, photo generated carriers could flow towards narrower bandgap well and be confined there before recombination taking place in SiGeSn barrier, and therefore enhances the QW emission. At temperature below 150 K, the intensity of QW peak increases dramatically due to further improved carrier confinement. Since majority carriers are generated in SiGeSn top barrier and then almost all of which are confined in direct bandgap GeSn well, the QW emission dominates the PL. At 10 K, the relatively weak peak at ~2130 nm is assigned to transition ③, and emissions from top barrier are almost invisible.

Temperature-dependent PL spectra using 1550-nm pumping laser is shown in Fig. 4(a). At temperatures from 300 to 200 K, the spectra are similar to those using 532-nm laser, i.e., a broad peak at 2000-2400 nm was observed at 300 K and a peak at ~2500 nm appears at 200 K. The clear difference can be seen from the PL spectra below 150 K: other than QW peak, a shorter wavelength peak at ~2150 nm was obtained, and its intensity significantly increases as temperature decreases to 20 K with the "growth rate" being comparable with that of QW peak. As a result, two peaks dominate the PL below 50 K.

The stacked temperature-dependent PL spectra were plotted to analyze the transitions, as show in Fig. 4(b). At 300 and 250 K, the broad peaks were fitted using the same method abovementioned and are detailed in Table III. Since the penetration depth of 1550-nm laser is much deeper than that of 532-nm laser, the light absorptions and emissions take place in all layers. At 300 K, three fitted peaks were obtained. The shorter wavelength peak (peak 1) consists of L to LH and L to HH transitions in both SiGeSn top and bottom barriers (①② and ⑨⑩), which is very close to the fitted shorter wavelength peak under 532-nm laser at 300 K. However, using 532-nm laser the transitions occur only in SiGeSn top barrier, while using 1550-nm the transitions in SiGeSn bottom barrier are also involved, leading to the broader peak full width at half maximum (FWHM) due to the slight differences of transition energies between SiGeSn top and bottom barriers.

The middle peak (peak 2) consists two groups of transitions: i) CB including $\Gamma$ and L valleys in QW to LH in both SiGeSn top and bottom barriers (③④⑦⑧); and ii) $\Gamma$ valley to HH and LH in GeSn buffer (⑪⑫). Due to close transition energy at 300 K these transitions cannot be further identified. The longer wavelength peak (peak 3) is clearly from QW emissions, including the first energy levels in $\Gamma$ valley to HH and L valley to HH transitions (⑤⑥). The intensity of this peak is relatively weak, which is explained as following: i) the QW layer is thin compared to other layers, and therefore the total absorption in QW is relatively weak; ii) at 300 K the carrier confinement is insufficient so that the photo-generated carriers could be thermally excited to populate L valley and then leaked out to barriers; and iii) the photo-generated carriers in barrier could recombine before their being collected by QW due to short lifetime.

As temperature decreases from 300 to 200 K, the intensity of peak 1 decreases due to the reduced phonon-assistance of indirect bandgap transitions. The intensity of peak 3 shows clear increase, which is attributed to the improved carrier confinement in QW. For peak 2, since the Ge buffer/GeSn buffer/SiGeSn bottom barrier forms a double heterostructure, and the GeSn buffer features direct bandgap, the emission from GeSn should be enhanced due to improved carrier confinement. On the other hand, the transitions from CB in QW to LH in SiGeSn barriers might be reduced as a result of carriers flowing towards the narrow bandgap GeSn well and buffer. Therefore, the overall intensity of peak 2 could be decreased. The intensity decrease of peak 1 and 2 leads to overall slightly decreased intensity of the broad peak at 2000-2400 nm, as can be seen in Fig. 4(a) inset.

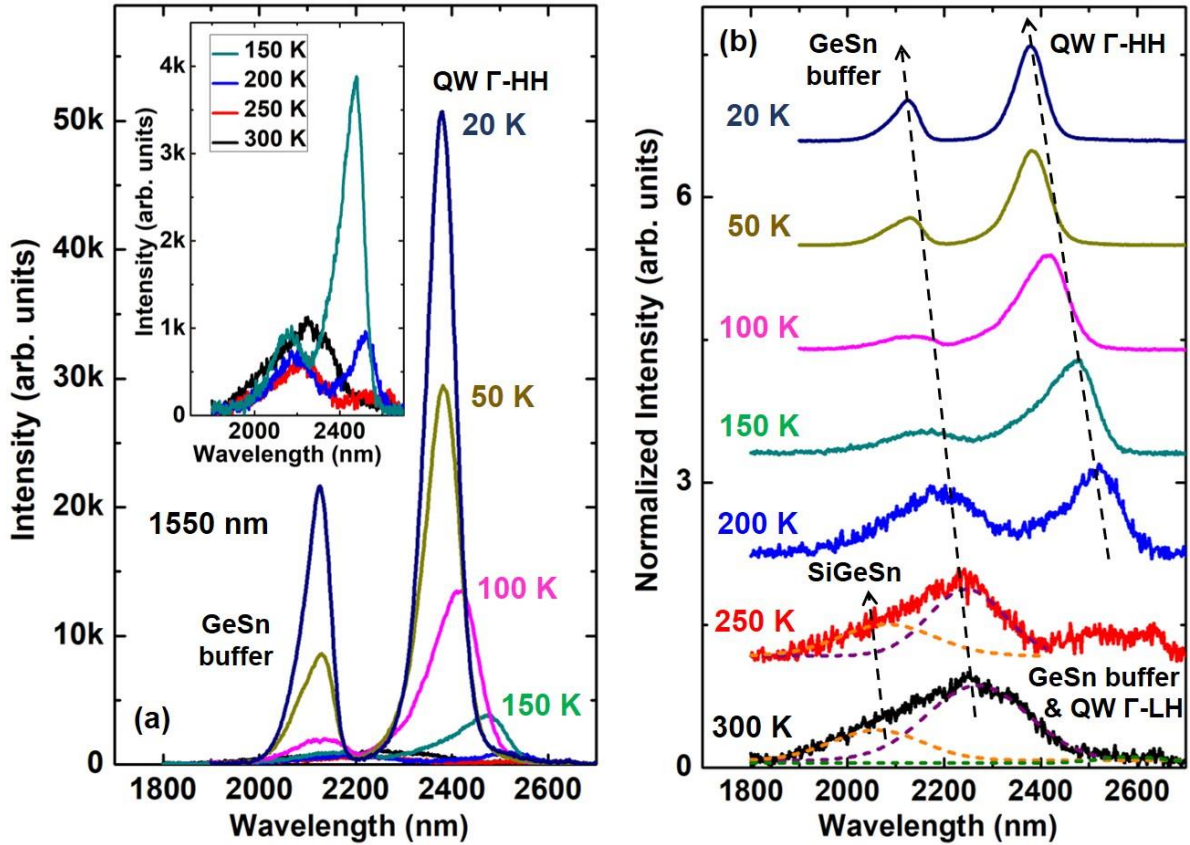

Figure 4 (a) Temperature-dependent PL under 1550 nm CW pumping laser. (b) Stacked normalized PL spectra showing the peak fitting.

As temperature further decreases, the direct bandgap transitions in QW (⑤) and GeSn buffer (⑪⑫) are dramatically enhanced due to further improved carrier confinement, while other transitions can be negligible. It can be seen that the intensity of GeSn buffer peak is almost the half of that of QW peak, indicating the considerable carrier collection efficiency in GeSn buffer.

The origins of carriers in GeSn buffer could be: i) direct absorption in GeSn buffer layer; and ii) carriers flowing from Ge buffer and SiGeSn bottom barrier layers.

Table III Summary of peak fitting under CW pumping lasers at 300 K

| Pumping Laser | 532 nm | | 1550 nm | | |
|---|---|---|---|---|---|
| PL at 300 K | 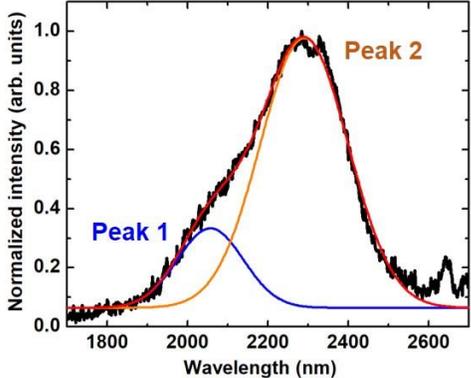 | | 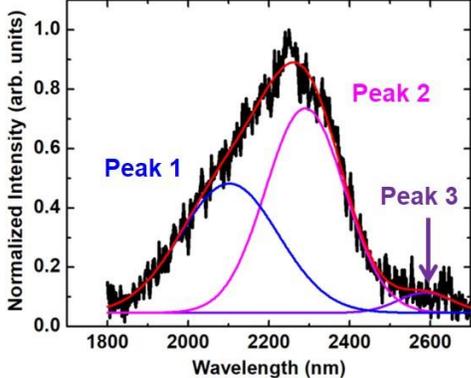 | | |
| Peak # | Peak 1 | Peak 2 | Peak 1 | Peak 2 | Peak 3 |
| Possible Transitions | ①② | ③④ | ①②⑨⑩ | ③④⑦⑧⑪⑫ | ⑤⑥ |
| Calculated peak positions from band diagram | ① 2147 nm ② 2039 nm | ③ 2259 nm ④ 2178 nm | 2023-2147 nm | 2175-2313 nm | ⑤ 2536 nm ⑥ 2446 nm |
| Fitted peak positions from PL spectra | 2056 nm | 2290 nm | 2101 nm | 2290 nm | 2587 nm |
| FWHM | 48 meV | 53 meV | 68 meV | 46 meV | 25 meV |

To further analyze the transitions, the pumping power-dependent PL study was conducted using the same setup configuration at 20 K. Figure 5(a) shows PL spectra under 532-nm pumping laser. The spectra were plotted using log-scale to clearly show the peaks. At very low pumping power (2 mW), the QW Γ to HH transition (⑤) dominates and the CB in QW to LH in SiGeSn top barrier transitions (③④) are almost invisible. As the pumping power increases, the Γ to HH transition peak increases significantly and the CB in QW to LH in SiGeSn transitions become more pronounced. Moreover, as the pumping power increases, the QW Γ to HH peak exhibits clear bule shift, which is due to the typical band filling effect in QW. While for the other peak, the peak position is almost unchanged, suggesting that this peak is originally from a transition in type-II band alignment, i.e., between CB in GeSn well and LH in SiGeSn top barrier.

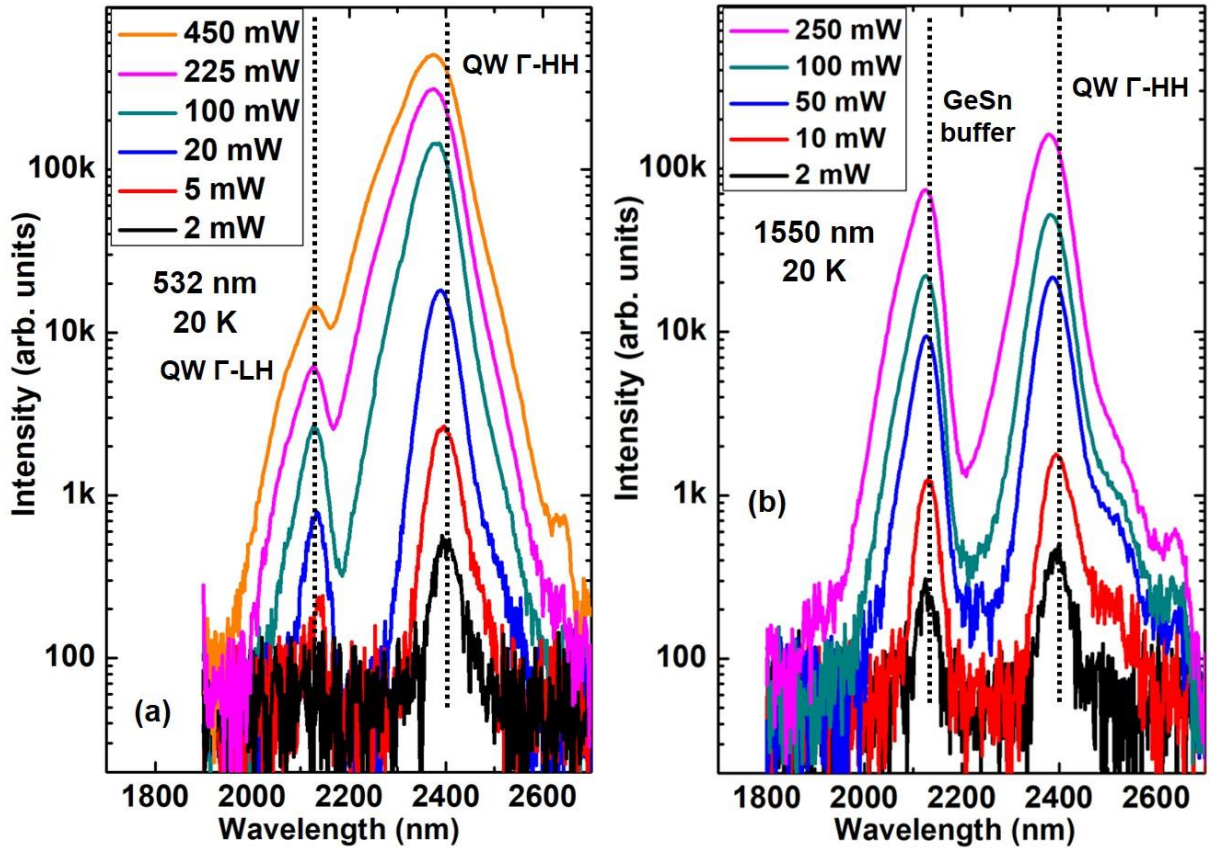

Figure 5 Pumping power-dependent PL spectra at 20 K under (a) 532-nm and (b) 1550-nm CW pumping lasers. The spectra were plotted using log-scale to clearly show the peaks.

The pumping power-dependent PL spectra under 1550-nm pumping laser at 20 K are shown in Fig. 5(b). As pumping power increases, both peaks increase with similar "growth rate", indicating a similar transition nature: both are from direct bandgap transitions. The longer wavelength peak that corresponds to QW Γ to HH transition (⑤) shows peak blue shift at higher pumping power as expected. While the shorter wavelength peak that corresponds to transitions from GeSn buffer (⑪⑫) shows a slight blue shift, which may be due to the thick GeSn buffer layer of 600 nm.

Note that for the GeSn buffer peak at higher pumping power that are greater than 100 mW, the asymmetric peak with broadened higher energy side was obtained. This indicates that due to the significantly increased number of photo generated carriers, the holes populated on SiGeSn barriers could partially recombine before their being collected by GeSn well or buffer, leading to the emission from CB in QW to LH in SiGeSn barriers transitions (③⑦) being observable.

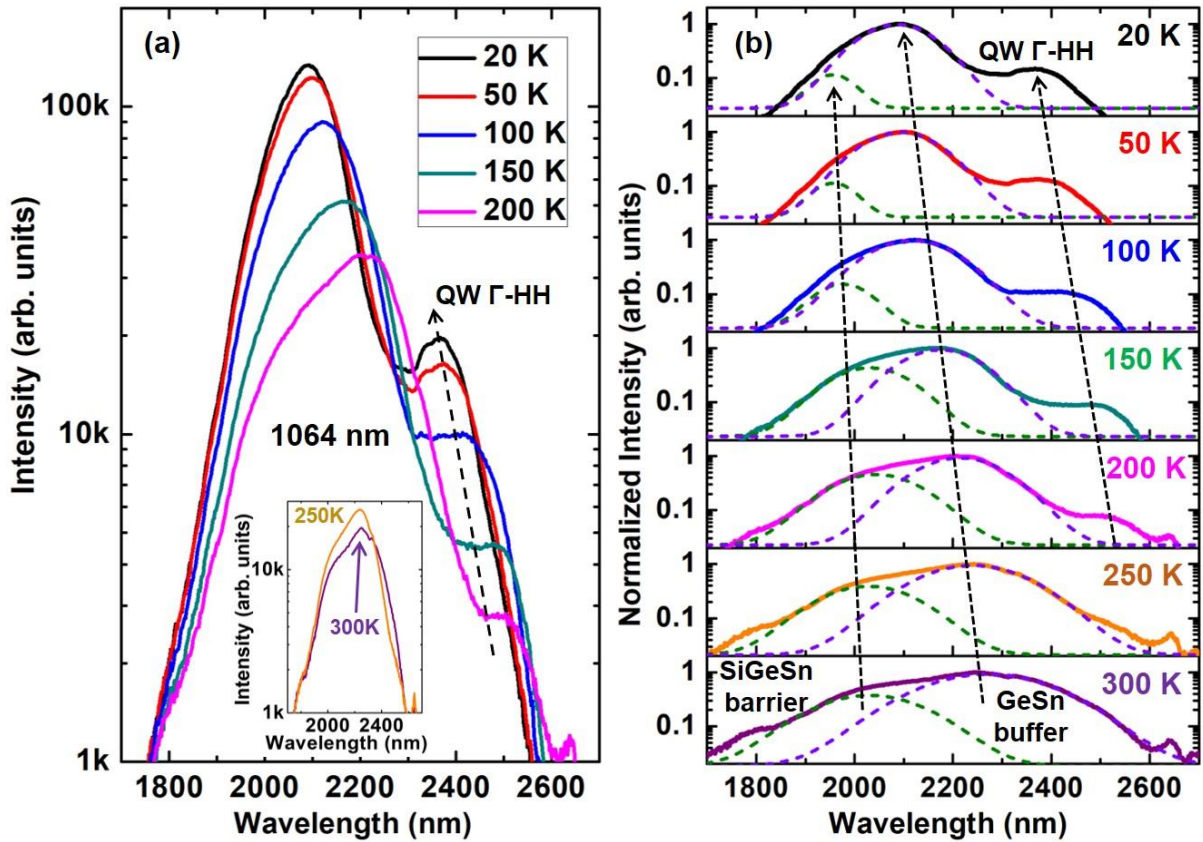

Figure 6 (a) Temperature-dependent PL under 1064 nm pulsed pumping laser. (b) Stacked normalized PL spectra showing the peak fitting. The spectra were plotted using log-scale.

To further understand the QW characteristics, PL behavior under pulsed lasers was investigated. Figure 6(a) shows the temperature-dependent PL spectra using 1064-nm pulsed pumping laser. The spectra were plotted using log-scale aiming to clearly show each peak. At 300 K, a broad peak was obtained which is similar to that under 1550-nm pumping laser. Considering the deep penetration depth of 1064-nm laser, emissions from all layers could contribute to the broad peak. As temperature decreases, the peak intensity increases as expected. At 200 K, a peak at ~2500 nm emerges, and this peak grows more pronounced as temperature decreases to 20 K. Unlike the PL spectra under CW laser, due to the intensity pumping using pulsed laser, the intensity of shorter wavelength broad peak is always much stronger than that of longer wavelength peak, i.e., $\Gamma$-HH transition in GeSn well. This is due to that the high photon injection level excites a huge number of photo-generated carriers that populates each layer simultaneously. The recombination could occur before the carriers re-distribute to the GeSn well via the carrier confinement effect even with the relatively long carrier lifetime at lower temperature.

Figure 6(b) shows the stacked temperature-dependent PL spectra with peak fitting using Gaussian distribution. The broad peak at 300 K was fitted with two peaks: a shorter wavelength peak at ~2050 nm corresponding to SiGeSn barriers transitions, and a longer wavelength peak at ~2300 nm corresponding to GeSn buffer transitions and from CB in QW to LH in SiGeSn barriers transitions. As temperature decreases, both peak intensities increase dramatically with the longer wavelength peak grows more rapidly than shorter wavelength peak, which is due to the improved carrier confinement in GeSn buffer as well as its the direct bandgap transition characteristic. On the other hand, considering the light absorption, the much thicker SiGeSn barrier layers and GeSn buffer layer lead to much more photo generated carriers compared to those in GeSn well, and under intense pulsed laser pumping the recombination occurs before carrier flowing to GeSn well. Therefore, although the carrier confinement in QW is sufficient at lower temperature, the relatively weak intensity of QW peak was obtained.

To confirm the peak origin, the PL comparison at 20 K between 1550-nm CW and 1064-nm pulsed lasers is plotted in Fig. 7. The fitted peaks under 1064-nm pumping laser are also plotted to facilitate the comparison. The fitted peak at ~2350 nm is close to QW peak under 1550-nm laser at ~2380 nm. The shorter wavelength of fitted peak could be due to the band filling that leads to the peak blue shift. Such peak blue shift was also observed for the fitted peak at ~2090 nm being compared with GeSn buffer peak under 1550-nm laser at ~2110 nm, and therefore confirm the origin of fitted peak at ~2090 nm. The fitted peak at ~1950 nm is assigned to transitions in SiGeSn barriers according to transition energies that are shown in Table II.

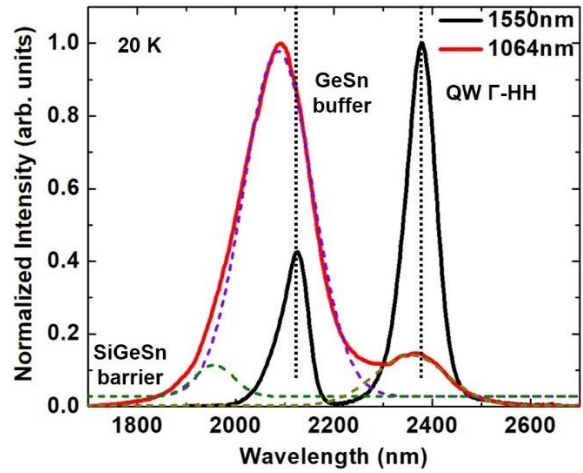

Figure 7 Comparison of PL at 20 K under 1550-nm CW and 1064-nm pulsed pumping lasers. The fitted peaks (dashed curves) under 1064-nm pumping laser are also shown.

To further confirm the GeSn well emission, a 1950-nm pulsed pumping laser with a long pass filter (pass for 2200 nm and longer wavelength) were used. The photon injection rate was tuned to match the 1064-nm laser. The temperature-dependent PL spectra are shown in Fig. 8. The irregular peaks at 300 and 250 K are due to the partially filtered out peak at the higher energy end. At 200 K, the emission from GeSn well can be clearly observed at ~ 2500 nm. As temperature further decreases, this peak grows significantly and shows peak blue shift as expected.

The PL spectra comparison (at 20 and 50 K) between 1064-nm and 1950 nm pulsed pumping lasers is plotted in Fig. 8 inset. It is clear that at each temperature, the position of the peak obtained from using 1950-nm laser matches well with that of longer wavelength peak using 1064-nm laser, unambiguously confirming that this peak is attributed to the QW emission.

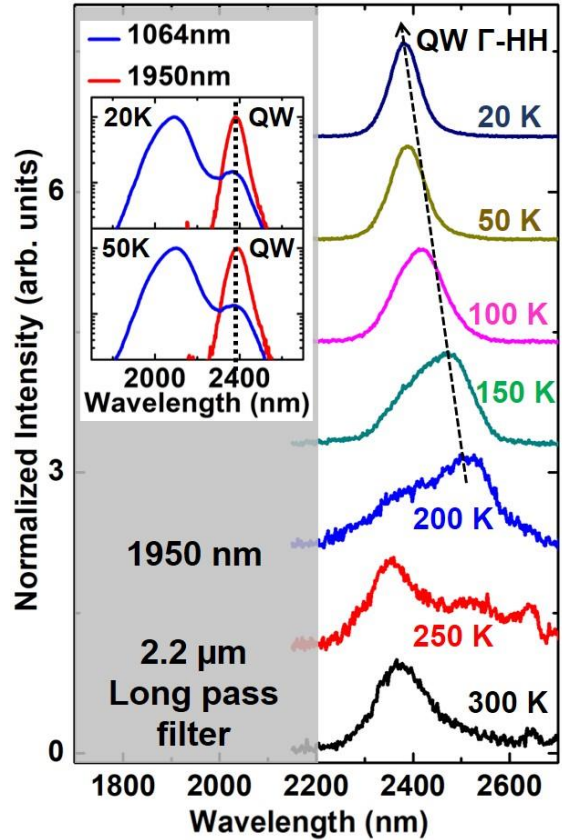

Figure 8 Temperature-dependent PL using 1950-nm pulsed pumping laser. A 2.2 µm long pass filter was used. Inset: comparison of PL spectra at 20 K and 50 K under 1064-nm and 1950-nm pulsed pumping lasers.

**V. DISCUSSION**

The absorption in SiGeSn barriers (both top and bottom) leads to the most photo-generated electrons populate the L-valleys, since the Γ-valleys are more than 50 meV above, and therefore very few electrons could jump to Γ-valley via thermal excitation at higher temperatures. This is confirmed by PL spectra that no peaks at ~1800 nm and 1900 nm were observed at 300 K, which correspond to Γ-HH and Γ-LH transitions, respectively. As a result, the optical transitions in SiGeSn barriers are dominated by L-HH and L-LH, as the HH-LH energy split is ~30 meV so that photo-generated carriers could populate both bands.

The GeSn buffer effectively eases the compressive strain in GeSn QW and ensures its bandgap directness. On the other hand, the presence of GeSn buffer alters the optical transition mechanism. The band diagram calculation indicates that the GeSn buffer features direct bandgap, and it also features narrower bandgap with respect to SiGeSn bottom barrier and Ge buffer, making photo generated carriers in SiGeSn bottom barrier could flow to GeSn buffer, which can be considered as carrier loss from the GeSn well. Although at this stage it is difficult to quantitatively estimate the carrier collection efficiency of GeSn buffer, however, since this "collection competition" only occurs at the barrier layer that is adjacent to GeSn buffer, for the devices using MQW, this type of carrier loss will not significantly deteriorate the carrier confinement in QW layers.

According to band diagram calculation, to lower the first energy level in Γ valley in CB so as to increase the Γ-L energy separation, thus enabling the enhancement of the carrier confinement in Γ valley, a thicker GeSn well could be grown. Moreover, using SiGeSn as barrier improves the barrier heights in Γ-valley and HH band compared to using GeSn, as shown in Table I. However, due to the reduced lattice constant because of Si incorporation, the SiGeSn barrier experiences tensile strain, which lifts LH band above the HH band. Since GeSn well features compressive strain, the type-II band alignment is then obtained in LH band. As the Γ-HH transition in GeSn well is preferred, such type-II band alignment results in a hole leakage channel at LH, which cannot be compensated by increasing the carrier injection level. In fact, a higher carrier collection efficiency in QW would occur at relatively lower injection level, at which the carriers populated on LH in SiGeSn barrier could flow to HH in GeSn well before recombination.

Theoretically, the Si and Sn compositions in SiGeSn ternary alloy can be selected at will to engineer the bandgap and lattice constant individually, allowing for nearly lattice matched material growth as well as providing sufficient barrier heights at both CB and VB. However, based on current material growth capability, the available Si and Sn compositions are limited. Therefore, the viable QW design to achieve type-I band alignment in each band relies on the progress of material growth technique. Additionally, to enhance the QW bandgap directness, increasing Sn composition in GeSn well is desirable. However, since GeSn well is grown pseudomorphically to GeSn buffer, the increase of Sn would induce higher degree of compressive strain of GeSn well, and thus push the bandgap towards indirectness. As a result, the Sn composition difference between GeSn well and buffer would have an optimized value, which might be close to 6% based on our experimental results.

It has been reported that during the material growth, the incorporation of Sn exhibits strongly strain sensitive to the starting growth surface, i.e., the compressive strain of the newly grown layer impairs the Sn incorporation, resulting in that the actual Sn composition being lower than the target.[15,33] As a result, for current devices using MQW, the slight variation of each individual well leads to the ununiformed injection and the quasi Fermi level would not perfectly lineup across all wells, and consequently deteriorates the device performance. To accurately grow MQW structure with the identical wells, the strain-balanced structures that were employed in III-V material system can be transplanted to GeSn and SiGeSn well-barrier system.[34-36] For the alternate tensile and compressively strained GeSn well and SiGeSn barrier layers, the zero average in-plane stress can be achieved by carefully design the thickness of each layer and selecting available Si and Sn composition in SiGeSn.

To improve the optical confinement, i.e., to increase the overlap between the optical field and the QW region, and to minimize the metal contact (usually at top surface) induced optical loss, the asymmetry structure can be considered, i.e., the thickness of the very top barrier needs to be optimized. In our previously reported GeSn MQW laser, the 70-80 nm SiGeSn cap layer was used, resulting in overall optical confinement factor of less than 10% in GeSn wells.[26] A thicker cap layer of several hundred nanometers would be suitable for SiGeSn/GeSn QW design. Further improvement of carrier and optical confinements can be realized by using separate confinement heterostructure (SCH). Note that even direct bandgap QW with type-I band alignment is attainable using desired SiGeSn barrier, the SCH would significantly improve the device performance. Due to the limit of available SiGeSn, the graded index cladding layers may be difficult to be obtained, and therefore the thicknesses of cladding layers need to be carefully designed to allow maximum overlap between optical field and QW region.

## VI. CONCLUSION

In summary, a SiGeSn/GeSn SQW was grown and used to comprehensively investigate the optical characteristics. Two CW lasers and two pulsed lasers were used as pumping sources to study the optical transitions. By analyzing the temperature- and pumping power-dependent PL spectra under different pumping lasers, with facilitation from band diagram calculation, the transitions from each layer were clearly identified and the transition mechanism was discussed in detail. By using GeSn buffer and SiGeSn barrier, the GeSn well features direct bandgap, and the improved barrier heights were achieved compared to our previous work. On the other hand, the type-II band alignment is obtained in LH due to tensile strain of SiGeSn barrier, which creates a hole leakage channel. As a result, the carrier injection level needs to be limited to ensure the QW carrier collection efficiency.

The revealed optical properties of QW in this work could provide a clear guidance for the future SiGeSn/GeSn based QW design.

## ACKNOWLEDGEMENT

The work is supported by the Air Force Office of Scientific Research (AFOSR) (Grant No.: FA9550-19-1-0341, FA9550-18-1-0045). Dr. Wei Du appreciates support from Provost's Research & Scholarship Fund at Wilkes University.

## DATA AVAILABILITY STATEMENT

The data that supports the findings of this study are available within the article.

## REFERENCE

1. R. A. Soref, Nat. Photo. **4**, 495-497 (2010).
2. R. Soref, D. Buca, and S.-Q. Yu, Optics and Photonics News **27**, 32-39 (2016).
3. A. W. Fang, H. Park, O. Cohen, R. Jones, M. J. Paniccia, and J. E. Bowers, Opt. Express **14**, 9203–9210 (2006).
4. H. Liu, T. Wang, Q. Jiang, R. Hogg, F. Tutu, F. Pozzi, and A. Seeds, Nature Photon. **5**, 416–419 (2011).
5. H. Yang, D. Zhao, S. Chuwongin, J. -H. Seo, W. Yang, Y. Shuai, J. Berggren, M. Hammar, Z. Ma, and W. Zhou, Nature Photon. **6**, 617–622 (2012).
6. K. Alberi, J. Blacksberg, L. D. Bell, S. Nikzad, K. M. Yu, O. D. Dubon, and W. Walukievicz, Phys. Rev. B **77**, 073202 (2008).
7. V. R. D'Costa, C. S. Cook, A. G. Birdwell, C. L. Littler, M. Canonico, S. Zollner, J. Kouvetakis, and J. Menéndez, Phys. Rev. B **73**, 125207 (2006).
8. S. A. Ghetmiri, W. Du, B. R. Conley, A. Mosleh, A. Nazzal, G. Sun, R. A. Soref, J. Margetis, J. Tolle, H. A. Naseem, and S. -Q. Yu, J. Vac. Sci. Technol., B **32**, 060601 (2014).
9. A. Mosleh, M. a. Alher, L.C. Cousar, W. Du, S.A. Ghetmiri, T. Pham, J.M. Grant, G. Sun, R. a. Soref, B. Li, H. a. Naseem, and S.-Q. Yu, Front. Mater. **2** (30), 1 (2015).
10. A. Mosleh, M. Alher, L.C. Cousar, W. Du, S.A. Ghetmiri, S. Al-Kabi, W. Dou, P.C. Grant, G. Sun, R.A. Soref, B. Li, H.A. Naseem, and S.Q. Yu, J. Electron. Mater. **45** (4), 2051-2058 (2016).

[11] J. Margetis, S.-Q. Yu, N. Bhargava, B. Li, W. Du, and J. Tolle, Semicond. Sci. Technol. **32** (12), 124006 (2017).

[12] J. Margetis, A. Mosleh, S. Al-Kabi, S.A. Ghetmiri, W. Du, W. Dou, M. Benamara, B. Li, M. Mortazavi, H.A. Naseem, S.Q. Yu, and J. Tolle, J. Cryst. Growth **463**, 128 (2017).

[13] S. A. Ghetmiri, W. Du, J. Margetis, A. Mosleh, L. Cousar, B. R. Conley, L. Domulevicz, A. Nazzal, G. Sun, R. Soref, J. Tolle, B. Li, H. Naseem, and S. Q. Yu, Appl. Phys. Lett. **105**, 151109 (2014).

[14] S. Wirths, R. Geiger, N. von den Driesch, G. Mussler, T. Stoica, S.Mantl, Z. Ikonic, M. Luysberg, S. Chiussi, J. M. Hartmann, H. Sigg, J. Faist, D. Buca, and D. Grützmacher, Nature Photon. **9**, 88–92 (2015).

[15] J. Margetis, S. Al-Kabi, W. Du, W. Dou, Y. Zhou, T. Pham, P. Grant, S. Ghetmiri, A. Mosleh, B. Li, J. Liu, G. Sun, R. Soref, J. Tolle, M. Mortazavi, and S. Yu, ACS Photonics **5**, 827 (2017).

[16] Wei Dou, Yiyin Zhou, Joe Margetis, Seyed Amir Ghetmiri, Sattar Al-Kabi, Wei Du, Jifeng Liu, Greg Sun, Richard Soref, John Tolle, Baohua Li, Mansour Mortazavi, and Shui-Qing Yu, Opt. Lett. **43**, 4558-4561 (2018).

[17] Y. Zhou, W. Dou, W. Du, S. Ojo, H. Tran, S. Ghetmiri, J. Liu, G. Sun, R. Soref, J. Margetis, J. Tolle, B. Li, Z. Chen, M. Mortazavi, and S. Yu, ACS Photonics **6**, 1434 (2019).

[18] J. Chrétien, N. Pauc, F. Armand Pilon, M. Bertrand, Q. Thai, L. Casiez, N. Bernier, H. Dansas, P. Gergaud, E. Delamadeleine, R. Khazaka, H. Sigg, J. Faist, A. Chelnokov, V. Reboud, J. Hartmann, and V. Calvo, ACS Photonics **6**, 2462 (2019).

[19] A. Elbaz, D. Buca, N. von den Driesch, K. Pantzas, G. Patriarche, N. Zerounian, E. Herth, X. Checoury, S. Sauvage, I. Sagnes, A. Foti, R. Ossikovski, J. Hartmann, F. Boeuf, Z. Ikonic, P. Boucaud, D. Grützmacher, and M. El Kurdi, Nature Photon. **14**, 375 (2020).

[20] Yiyin Zhou, Yuanhao Miao, Solomon Ojo, Huong Tran, Grey Abernathy, Joshua M. Grant, Sylvester Amoah, Gregory Salamo, Wei Du, Jifeng Liu, Joe Margetis, John Tolle, Yong-hang Zhang, Greg Sun, Richard A. Soref, Baohua Li, and Shui-Qing Yu, Optica **8**, 924-928 (2020).

[21] Wei Dou, Seyed Ghetmiri, Sattar Al-Kabi, Aboozar Mosleh, Yiyin Zhou, Bader Alharthi, Wei Du, Joe Margetis, John Tolle, Andrian Kuchuk, Mourad Benamara, Baohua Li, Hameed Naseem, Mansour Mortazavi, and Shui-Qing Yu, J. Electron. Mater. **45**, 6265-6272 (2016).

[22] S. A. Ghetmiri, Y. Zhou, J. Margetis, S. Al-Kabi, W. Dou, A. Mosleh, W. Du, A. Kuchuk, J. Liu, G. Sun, R. A. Soref, J. Tolle, H. A Naseem, B. Li, M. Mortazavi, S.-Q. Yu, Opt. Lett. **42**, 387-390 (2017).

[23] Wei Du, Seyed Amir Ghetmiri, Joe Margetis, Sattar Al-Kabi, Yiyin Zhou, Jifeng Liu, Greg Sun, Richard A. Soref, John Tolle, Baohua Li, Mansour Mortazavi, and Shui-Qing Yu, Journal of Applied Physics **122**, 123102 (2017).

[24] Perry Grant, Joe Margetis, Yiyin Zhou, Wei Dou, Grey Abernathy, Wei Du, Baohua Li, John Tolle, Jifeng Liu, Greg Sun, Richard A. Soref, Mansour Mortazavi, Shui-Qing Yu, AIP Advances **8**, 025104 (2018).

[25] Perry Grant, Joe Margetis, Wei Du, Yiyin Zhou, Wei Dou, Grey Abernathy, Andrian Kuchuk, Baohua Li, John Tolle, Jifeng Liu, Greg Sun, Richard Soref, Mansour Mortazavi and Shui-Qing Yu, Nanotechnology **29**, 4652 (2018).

26  Joe Margetis, Yiyin Zhou, Wei Dou, Perry C. Grant, Wei Du, Bader Alharthi, Huong Tran, Solomon Ojo, Grey Abernathy, Seyed A. Ghetmiri, Jifeng Liu, Greg Sun, Richard Soref, John Tolle, Baohua Li, Mansour Mortazavi, Shui-Qing Yu, Appl. Phys. Lett. **113**, 221104 (2018).

27  D. Stange, N. Von Den Driesch, D. Rainko, C. Schulte-Braucks, S. Wirths, G. Mussler, A. T. Tiedemann, T. Stoica, J. M. Hartmann, Z. Ikonic, S. Mantl, D. Grutzmacher, D. Buca, Opt. Express, **24**, 1358-1367 (2016).

28  Stange D, von den Driesch N, Rainko D, Roesgaard S, Povstugar I, Hartmann JM, Stoica T, Ikonic Z, Mantl S, Grützmacher D, Buca D. Optica, **4**, 185-188 (2017).

29  Nils von den Driesch, Daniela Stange, Denis Rainko, Ivan Povstugar, Peter Zaumseil, Giovanni Capellini, Thomas Schröder, Thibaud Denneulin, Zoran Ikonic, Jean-Michel Hartmann, Hans Sigg, Siegfried Mantl, Detlev Grützmacher, and Dan Buca, Adv. Sci. **5**, 1700955 (2018).

30  D. Stange, N. von den Driesch, T. Zabel, F. Armand-Pilon, D. Rainko, B. Marzban, P. Zaumseil, J. Hartmann, Z. Ikonic, G. Capellini, S. Mantl, H. Sigg, J. Witzens, D. Grützmacher, and D.Buca, ACS Photonics **5**, 4628 (2018).

31  N. von den Driesch, D. Stange, D. Rainko, U. Breuer, G. Capellini, J. M. Hartmann, H. Sigg, S. Mantl, D. Grutzmacher, and D. Buca, Solid State Electronics **155**, 139-143 (2019).

32  J. Margetis, S. Ghetmiri, W. Du, B. Conley, A. Mosleh, R. Soref, G. Sun, L. Domulevicz, H. Naseem, S. Yu, J. Tolle, ECS Trans. **64**, 711-720 (2014).

33  Wei Dou, Mourad Benamara, Aboozar Mosleh, Joe Margetis, Perry Grant, Yiyin Zhou, Sattar Al-Kabi, Wei Du, John Tolle, Baohua Li, Mansour Mortazavi, Shui-Qing Yu, Scientific Reports **8**, 1-11 (2018).

34  N. J. Ekins-Daukes, K. Kawaguchi, and J. Zhang, Crystal Growth & Design **2**, 287-292 (2002).

35  V. D. Jovanović, Z. Ikonić, D. Indjin, and P. Harrison, J. Appl. Phys. **93**, 3194 (2003).

36  Guo-En Chang, Shu-Wei Chang, and Shun Lien Chuang, IEEE J. Quantum Electron **46**, 1813-1820 (2010).